# Leveraging Optical Anisotropy of the Morpho Butterfly Wing for Quantitative, Stain-Free, and Contact-Free Assessment of Biological Tissue Microstructures

*Paula Kirya, Aida Mestre-Farrera, Jing Yang, Lisa V. Poulikakos\**

P. Kirya, L.V. Poulikakos
Department of Mechanical and Aerospace Engineering, Program of Materials Science and Engineering, UC San Diego, 9500 Gilman Drive, La Jolla, CA, 92093, USA
E-mail: lpoulikakos@ucsd.edu

A. Mestre-Farrera, J. Yang
Department of Pharmacology, Moores Cancer Center, UC San Diego, 3855 Health Sciences Drive, La Jolla, CA, 92093, USA
Department of Pediatrics, UC San Diego, 9500 Gilman Drive, La Jolla, CA, 92093, USA

Changes in the density and organization of fibrous biological tissues often accompany the progression of serious diseases ranging from fibrosis to neurodegenerative diseases, heart disease and cancer. However, challenges in cost, complexity, or precision faced by existing imaging methodologies pose barriers to elucidating the role of tissue microstructure in disease. Here, we leverage the intrinsic optical anisotropy of the Morpho butterfly wing and introduce Morpho-Enhanced Polarized Light Microscopy (MorE-PoL), a stain- and contact-free imaging platform which enhances and quantifies the birefringent material properties of fibrous biological tissues. We develop a mathematical model, based on Jones calculus, which quantifies fibrous tissue density and organization. As a representative example, we analyze collagen-dense and collagen-sparse human breast cancer tissue sections and leverage our technique to assess the microstructural properties of distinct regions of interest. We compare our results with conventional Hematoxylin and Eosin (H&E) staining procedures and second harmonic generation (SHG) microscopy for fibrillar collagen detection. Our findings demonstrate that our MorE-PoL technique provides a robust, quantitative, and accessible route toward analyzing biological tissue microstructures, with great potential for application to a broad range of biological materials.

# 1. Introduction

Progression of many common and fatal diseases including fibrosis, various cancers, neurodegenerative diseases, and heart disease is accompanied by changes and rearrangement of fibrous tissue microstructure.[1] A common byproduct of such diseases is characterized by excessive extracellular matrix (ECM) buildup. The ECM contains large quantities of supramolecular collagen fibers, which can form highly uniaxial assemblies.[2–5] When interacting with light, such collagen fiber arrangements exhibit birefringence, a form of optical anisotropy which arises due to varying refractive indices along orthogonal optical axes. Beyond collagen, versatile tissue and cellular components including nuclei,[6] elastin,[6] and amyloids[7] exhibit such optical anisotropy. However, naturally occurring birefringence in biological tissues is inherently weak, posing significant challenges toward its effective characterization with polarized light.[8] Existing methodologies to quantify biological tissue microstructure face trade-offs in precision and experimental complexity.[9,10] Subsequently, elucidating the role of tissue microstructures in the origin and progression of disease remains challenging.

The most common approach to histological tissue examination is the imaging of chemically stained tissue sections with light microscopy. Hematoxylin and Eosin (H&E) is the general gold standard for histopathological imaging, which stains nuclei blue, and connective tissue pink.[11] Masson's trichrome affords more specificity, as it stains collagen blue, nuclei dark brown, muscle tissue red, and cytoplasm pink.[12] These staining techniques provide largely qualitative, morphological information, some of which can be quantified with image analysis algorithms,[13,14] yet they do not directly quantify tissue microstructures.[11][16] Moreover, staining analysis must account for artifacts that arise from incomplete fixation, section thickness, staining variation or interpretation bias that may render inaccurate results.[15] Stain-free methods can significantly expedite tissue analysis by circumventing laborious preparation and fixation, thereby also enabling in-vivo applications. Moreover, by removing preparation steps required for staining that may generate erroneous artifacts, stain-free imaging can result in more accurate data, specific to the physical nature of the specimen.

High-precision stain-free methods are available to characterize tissue microstructure.[16] Polarized light microscopy, where a sample is assessed between crossed linear polarizers, can distinguish between isotropic and anisotropic materials.[17,18] This method can reveal information about the structure and composition of a material based on its unique optical anisotropy.[19] Many components of biological tissue – particularly where there are thin fibers – exhibit weak birefringence, e.g., in collagen, the change of refractive indices along orthogonal optical axes is Δn=0.003.[17,19] This poses challenges in the visualization of optical fluctuations corresponding to the optical anisotropy of tissue.[19] Polarization-sensitive stains, such as Picrosirius Red which binds to collagen[20] or Congo Red for amyloids[21,22] can enhance birefringence, yet they are subject to the aforementioned challenges and artifacts arising from staining. Moreover, a tissue's inherent autofluorescence is amplified when paraffin-embedded, which disrupts the signal produced e.g. with Picrosirius Red.[11]

Polarimetric imaging – such as Stokes, Mueller Matrix, and Jones polarimetry – measures the polarization state of light that is transmitted or reflected by a sample.[8,23–26] With these techniques, the optical effects corresponding to the tissue polarization can only be visualized after processing the image data. Second harmonic generation (SHG) microscopy allows for non-invasive, high-spatial-resolution (400-100 nm) visualization of tissue fibers (most commonly collagen).[5,27,28] Optical coherence tomography enables in-vivo imaging, performing a real-time optical biopsy and allowing for direct 3D histological visualization of human tissue, albeit at comparably lower spatial resolutions (20-5 μm).[29–32] Thus, existing stain-free methods for visualizing the tissue microstructure typically require expensive and complex equipment necessitating extensive expertise for their operation and analysis.

Photonic surfaces, where assemblies of optical micro- and nanostructures enable distinct interaction with light, hold great potential to miniaturize complex light-matter interactions and circumvent the need for prohibitively complex optical equipment.[33–38] In biological imaging and sensing, metasurfaces, which are composed of artificially engineered optical nanostructures, have demonstrated visualization and detection of optical and vibrational properties of biological molecules,[39–48] proteins,[40,49–52] extracellular vesicles,[53] viruses,[54,55] and cells.[56–60] Histological tissue sections have been interfaced with plasmonically active microscope slides that generate colors specific to the dielectric function of its surrounding material, allowing for the identification of cancerous epithelial cells in a stain-free imaging configuration.[61] Moreover, plasmonic metasurfaces that operate within the mid-infrared (MIR) spectrum were conceptualized for the qualitative chemical analysis of histological tissues.[58,62,63]

The abovementioned metasurface techniques achieve enhanced visualization of biological tissue by leveraging evanescent near fields, which decay exponentially perpendicular to the metasurface interface. The effectiveness of these approaches necessitates direct contact with the biological tissue, thus hindering probing of the entire depth of the histological section and limiting the repetitive use of the metasurfaces, which frequently require laborious fabrication procedures.[64] Photothermal heating, which can accompany the optical resonances of the nanostructures comprising metasurfaces, may damage the studied tissue.[65,66] The development of photonic surfaces, which visualize biological tissue in a contact-free manner, would help overcome these challenges. Moreover, while the abovementioned approaches leverage metasurfaces to assess optical and vibrational properties of biological media, they do not quantify their microstructure.

Here, we leverage the inherent, strong birefringence of the Blue Morpho butterfly wing, an optically anisotropic photonic crystal derived from nature, and introduce a quantitative, stain-free and contact-free imaging technique to probe the microstructural properties of histological tissue sections. Our imaging methodology, termed Morpho-Enhanced Polarized Light Microscopy (MorE-PoL), can be directly implemented in a commercially available polarized light microscope, circumventing the need for complex optical components and configurations. Moreover, the direct implementation of a nature-derived photonic surface does not require laborious fabrication procedures needed for artificial photonic surfaces, allowing for facile implementation of MorE-PoL, including in under-resourced settings. Based on our previous computational proof-of-concept study,[17,67] we develop a Jones-calculus model, allowing for the comparative, quantitative assessment of the density and organization of fibrous biological tissue. We demonstrate the potential of our imaging technique by analyzing collagen-dense and collagen-sparse human breast cancer tissue sections.

In nature, we observe organisms that can leverage photonics at the micro- and nanoscale in their prominent anatomy including wings, shells, leaves, and skins. Specific examples of such organisms include butterflies of the Morpho, Junoia, and Papilio genus, Jewel beetles, organisms of the Cephalopoda class (e.g. squid, cuttlefish, octopus), several plants including those of the genus Selaginellay, diatoms, the feathers of peacocks, hummingbirds, and the Eurasian Jay.[68–76] A prominent example of this feature is the Blue Morpho butterfly wing. An example species is the *Morpho menelaus*, whose dorsal surface is covered with a double layer of scales, comprised of parallel longitudinal striations with periodic separation between ridges.[77,78] Vertically stacked chitin lamellae that make up the ridges produce a multilayer interference effect that increases the reflectivity of the wing and the purity of the reflected color.[77] The diffraction-grating arrangement of its longitudinal ridges separates light into different wavelengths depending on incident angle.[77] This means hue is shifted when angles deviate from normal incidence or viewing angle.[77,79]

Another profound optical property of the Morpho wing is its strong ability to polarize light. The directionality of the longitudinal ridges derives a sensitivity to the transverse electric (TE) and transverse magnetic (TM) modes of light propagation where the electric and magnetic fields are parallel or perpendicular to the gratings, respectively.[77] The reflection spectra of *Morpho menelaus* are shown to shift from the blue regime when illuminated by a transverse electric (TE) wave towards green when illuminated with a transverse magnetic (TM) wave.[77] Further, when imaged between crossed polarizers the scales appear darker when the orientations of the ridges coincide with the polarizer or analyzer, but the scales are bright when oriented at 45°.[77,80] This means that the scales introduce a phase delay unto the light that produces a high degree of elliptical polarization.[17,24,77] Thus, the structural anisotropy of the Morpho wing enables mapping of the polarization state of light to unique optical responses.

**Figure 1** schematically illustrates the principle of Morpho-Enhanced Polarized Light Microscopy (MorE-PoL) introduced in this work. As a representative system of fibrous biological tissue, we study patient derived xenograft (PDX) models for treatment-naïve triple-negative breast cancer (TNBC) subtypes (Figure 1a). We select two tissue types based on stiffness, which exhibit dense or sparse collagen content, respectively. The paraffin-embedded samples were sectioned at 3 μm thickness (see section 4. Methods). Figure 1b shows a photograph and scanning electron micrographs of the studied Morpho butterfly wing, which is placed onto a glass microscope coverslip and covered with an additional glass coverslip to protect the scales upon imaging (see section 4. Methods). Our MorE-PoL experimental configuration is shown in Figure 1c, where the Morpho wing and tissue sections shown in parts a and b are placed on top of each other on the microscope stage. Importantly, this means that no direct contact between the Morpho wing and the tissue of interest is needed for this technique, as it relies on polarized light-matter interactions in the far field. The inset of Figure 1c shows schematically how linearly polarized light is incident on the tissue of interest. Upon traversing the fibrous tissue sample, the polarized light will gain a degree of ellipticity. The strong optical anisotropy of the Morpho wing will then further alter the ellipticity of the light in reflection before it passes through the tissue section a second time and enters an analyzer orthogonal to the orientation of incident linearly polarized light. Thus, this method enables the specific enhancement of the optical anisotropy of the fibrous tissue by leveraging the optical anisotropy of the Morpho wing. Subsequently, changes in the density, organization, and orientation of collagen fibers in the tissue of interest can be quantitatively assessed with MorE-PoL, while they would fall below detection thresholds if the Morpho wing were absent in conventional polarized light microscopy. Because this method is contact free and stain free, the Morpho wing can be reused over many analysis cycles for a multitude of different tissue sections, significantly accelerating and democratizing experimental procedures to assess biological tissue microstructure.

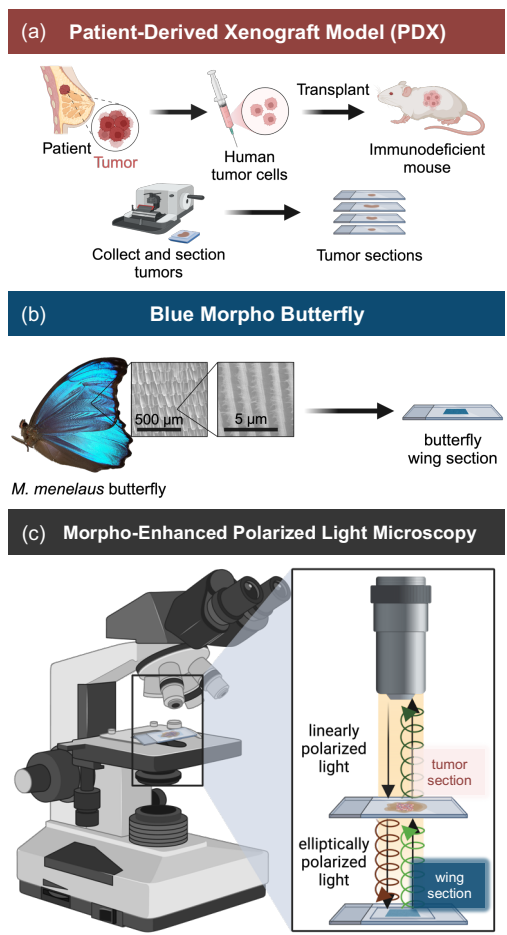

*Figure 1:* *Schematic of Morpho-Enhanced Polarized Light Microscopy (MorE-PoL), the imaging platform introduced in this work. (a) The studied histological murine human breast cancer xenograft tissue sections (3µm thickness) are derived from a Patient-Derived Xenograft Model (see section 4. Methods). (b) Optical anisotropy and structural color from the Morpho menelaus butterfly wing was leveraged in this work (see section 4. Methods). (c) Principle of operation of Morpho-Enhanced Polarized Light Microscopy. The coverslip containing the Morpho wing section was placed underneath the coverslip with the histological tumor section onto the rotating stage of the polarized light microscope. This arrangement was imaged between crossed linear polarizers and the stage was rotated counterclockwise from 0°-180° in 15° increments. Inset: Linearly polarized light transmitted through fibrous biological tissue obtains a degree of ellipticity which is further enhanced upon interaction with the Morpho butterfly wing. The output signal is imaged through a linear analyzer and assessed with Jones calculus. Schematic was generated in Biorender.*

## 2. Theoretical Model

Within an identified region of interest in an imaged tissue section, our MorE-PoL technique acquires the average pixel intensity at varying sample stage rotation angles. The resulting intensity profile of the image series was fit using a mathematical model described in this section. Based on Jones calculus, which assumes fully polarized light excitation and linear optical components, this model characterizes light propagation in optically anisotropic regions of the sample of interest. In the example case studied here, regions containing collagen in the tissue of interest exhibit optical anisotropy.

The Jones matrices of the horizontal (oriented along the x-axis) and vertical (oriented along the y-axis) polarizers, respectively are:

$$P_x = \begin{pmatrix} 1 & 0 \\ 0 & 0 \end{pmatrix} \text{ and } P_y = \begin{pmatrix} 0 & 0 \\ 0 & 1 \end{pmatrix}. \tag{1}$$

The Jones matrix of a wave retarder that is rotated by an angle $\psi$ in the counterclockwise direction and induces a phase shift $\delta$ along its optical axis between the two orthogonal polarizers is given by:

$$J_{M+T}(\delta, \psi) = \begin{pmatrix} \cos\psi & -\sin\psi \\ \sin\psi & \cos\psi \end{pmatrix} \times \begin{pmatrix} e^{i\delta/2} & 0 \\ 0 & e^{-i\delta/2} \end{pmatrix} \times \begin{pmatrix} \cos\psi & \sin\psi \\ -\sin\psi & \cos\psi \end{pmatrix}. \tag{2}$$

As the Morpho wing and the studied tissue section (M+T) are simultaneously rotated together in counterclockwise direction by the rotation angle $\theta$ between the two fixed orthogonal polarizers, the rotation of M+T is $\psi = \phi + \theta$, where $\phi$ is the optical axis orientation of the combined M+T system, described by the Jones matrix $J_{M+T}$.

The incident light passes through a linear polarizer oriented along x as $\mathbf{E}_x = P_x \cdot \mathbf{E}_0$, where $\mathbf{E}_0$ is the electric field vector of the light source. When plane-polarized light ($\mathbf{E}_x$) enters the Morpho + Tissue arrangement, represented by the combined Jones matrix $\mathbf{J}_{M+T}$, the reflected electric field is described by $\mathbf{E}_{M+T} = \mathbf{J}_{M+T}(\delta, \phi + \theta) \cdot \mathbf{E}_x$. Subsequently, $\mathbf{E}_{M+T}$ transmits through a vertically oriented analyzer ($P_y$) such that the output electric field captured by the detector is $\mathbf{E}_{out} = P_y \cdot \mathbf{E}_{M+T}$. Using $I \sim |\mathbf{E}_{out}|^2$, the output light intensity is calculated to produce the second-power sinusoidal intensity profile:

$$I(\theta) = I_0 \sin(2\phi + 2\theta)^2 \sin\left(\frac{\delta}{2}\right)^2, \tag{3}$$

where we employ $I_0 = 1$, equivalent to the normalized grayscale pixel value of the incident lamp intensity. The phase delay stemming from linear birefringence $\delta$ is given by:[81]

$$\delta \approx \frac{2\pi}{\lambda} \Delta n d \tag{4}$$

where $\lambda$ is the wavelength of incident light, $\Delta n$ is the local birefringence, and d is the tissue thickness. Previous studies have established that increased presence of collagen fibers corresponds to an increase in birefringence ($\Delta n$).[82] As the incident wavelength and sample thickness (3μm) are fixed parameters, the intensity profile equation provides a direct measure of the relative fiber orientation and fiber density within a region.

**Figure 2** illustrates our Jones calculus model (Equation 3) which quantifies the fibrous properties of biological tissue when interfaced with the Morpho butterfly wing. We divide our observations into 4 distinct cases, where the phase delay $\delta$ provides information on dense (↑ $\delta$, Figure 2a,b,e,f) vs. sparse (↓ $\delta$, Figure 2c,d,g,h) collagen density and the $R^2$ fit value of Equation 3 provides information on ordered (↑ $R^2$, Figure 2a,c,e,g) vs. disordered (↓ $R^2$, Figure 2b,d,g,h) collagen content. Note that realistic cases may lie between the extreme scenarios of dense vs. sparse and ordered vs. disordered collagen content, which were demonstrated in Figure 2 for clarity. Jones fit data for each case was simulated in MATLAB. $\delta = 90°$ and $\delta = 45°$ was inserted into the Jones calculus intensity profile for ↑ $\delta$ and ↓ $\delta$ cases, respectively. For ↑ $R^2$, $\phi = 0$ was inserted into the intensity profile. To obtain a profile with ↓ $R^2$, an array of 13 random $\phi$ values for $\phi \in [0°, 45°]$ was generated with the MATLAB randi function. This array was then inserted into the Jones fit equation to produce intensities for each 15° of rotation. Following the procedure we undergo with experimental data, the MATLAB Curve Fitter tool was implemented to estimate $\phi$ and $\delta$ values that produce the best fit of the raw data (see section 4. Methods). The non-linear least squares method and trust-region algorithm were selected, as well as the constraints $\phi \in [0°, 180°]$ and $\delta \in [0°, 180°]$. The $R^2$ value of the fit is generated by the curve-fitting tool.

Note that as the Jones fit for the intensity profile decreases in certainty with lower $R^2$ values, the range of possible $\delta$ values also increases. This is seen in the simulated Case 2, where we model an extreme case of density and disorder (Figure 2b,f). The Jones fit (Equation 3) yields $\delta = 48.048° \pm 76.524°$, despite the inserted pixel intensities having been derived for a $\delta = 90°$ value. When comparing the $\delta$ values of Cases 1 and 2, we cannot definitely ascertain one being denser than the other given the wide 95% confidence interval for Case 2.

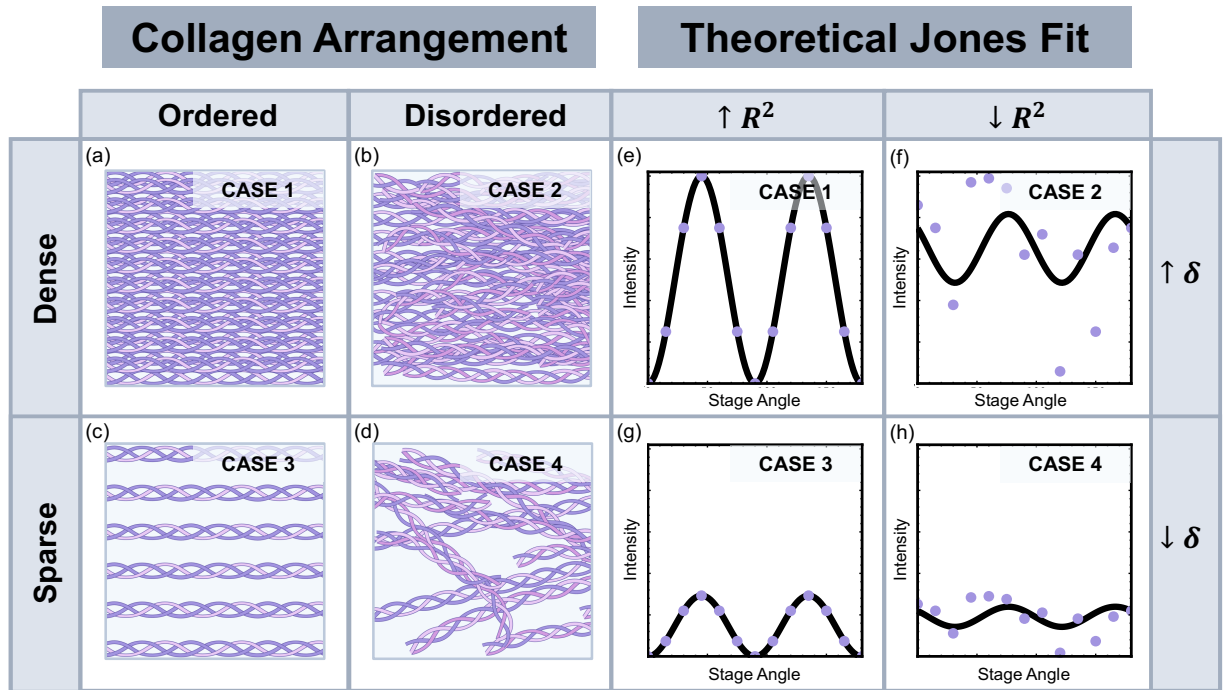

*Figure 2:* Table describing collagen arrangements (a-d) corresponding to distinct intensity profiles fit to our Jones calculus model (Equation 3) (e-h). (a,e) Case 1: ordered and dense, ↑ $R^2$, ↑ $\delta$, (b,f) Case 2: disordered and dense, ↓ $R^2$, ↑ $\delta$, (c,g) Case 3: ordered and sparse, ↑ $R^2$, ↓ $\delta$, (d,h) Case 4: disordered and sparse, ↓ $R^2$, ↓ $\delta$. (e) Case 1: $R^2 = 1.0$, $\delta = 90.00° \pm 0.063°$, (f) Case 2: $R^2 = 0.1818$, $\delta = 48.048° \pm 76.524°$, (g) Case 3: $R^2 = 1.0$, $\delta = 45.00° \pm 0.000°$, (h) Case 4: $R^2 = 0.1818$, $\delta = 25.46° \pm 38.778°$. Collagen Arrangement illustrations were generated in Biorender.

## 2. Experimental Results and Discussion

### 2.1. Characterization of Optical Anisotropy in the Morpho butterfly wing

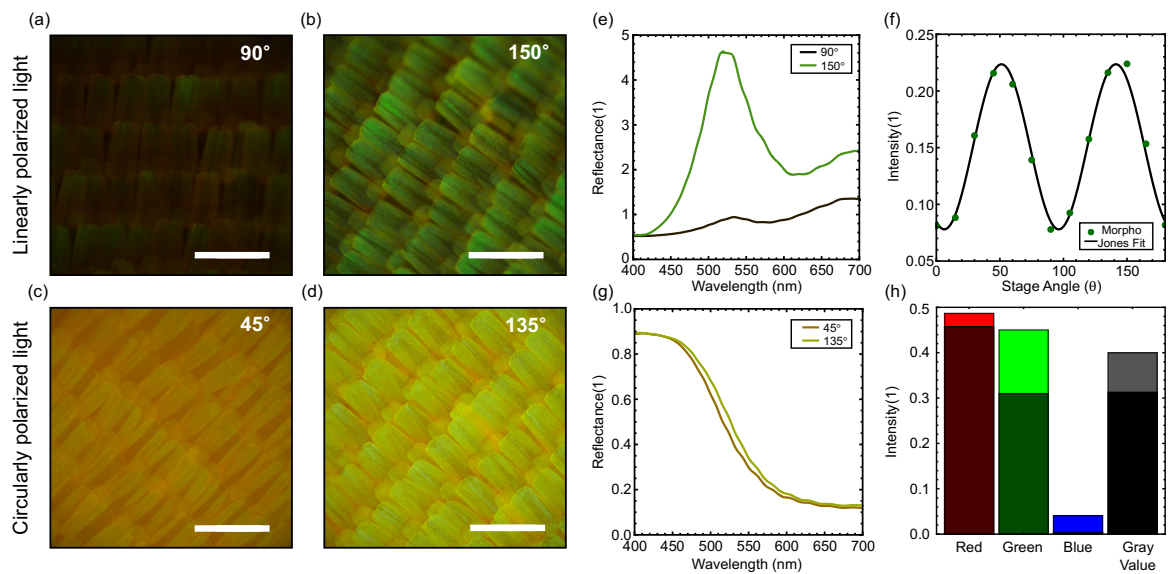

*Figure 3:* Experimental characterization of the Morpho butterfly wing. (a-d) Polarized light microscope images of the Morpho butterfly wing at varying optical axis orientations between crossed linear polarizers (parts a,b) and for circularly polarized light excitation with a linear analyzer (parts c,d). (e) Reflectance spectra of the Morpho butterfly wing between crossed

*linear polarizers at 90° and 150° stage orientations. (f) Jones calculus fit (Equation 3) of the Morpho butterfly wing between crossed linear polarizers at varying microscope stage angle orientations. (g) Reflectance spectra of the Morpho butterfly wing for circularly polarized light excitation at 45° and 135° stage orientations. (h) Histogram of the mean normalized Red, Green, Blue and Gray Value color signals obtained for images shown in parts c at 45° (darker histograms), and d at 135° (lighter histograms) stage orientations.*

**Figure 3** characterizes the anisotropic optical properties of the Morpho butterfly wing upon linearly and circularly polarized light excitation with a linear analyzer. The Morpho wing section was rotated counterclockwise from $\theta = 0°\text{-}180°$ in 15° increments. Two series of imaging took place, the first in which the sample was illuminated with horizontally polarized light and an orthogonal analyzer, and for the latter a quarter-wave retarder was introduced to the light path for circularly polarized light excitation and the identical linear analyzer.

Illumination and analysis with orthogonal linear polarizers resulted in the highest reflectance of the Morpho butterfly wing at $\theta = 150°$ and lowest at $\theta = 90°$ stage orientation (Figures 3a-d). For each image of the wing section, the pixel intensity was averaged and normalized by maximum pixel intensity (255) in MATLAB. The average intensity for each orientation was plotted and fitted to the Jones Calculus model (Equation 3) with MATLAB's built-in nonlinear least squares function in the Curve-Fitting tool (Figure 3f). The Jones calculus fit yielded a retardance ($\delta$) of 47.08°±3.707°. The high accuracy of the fit ($R^2$=0.9890) demonstrates that we can accurately model the linear behavior of the Morpho wing as a wave retarder.

Spectra of the Morpho wing at the recorded orientations producing its minimum ($\theta = 90°$) and maximum ($\theta = 150°$) intensities under linearly polarized light illumination are shown in Figure 3e. At $\theta = 150°$, the reflectance, normalized with respect to the background signal of crossed linear polarizers, peaks at 4.637 at 518.4 nm. At $\theta = 90°$, this reflectance is damped to 0.8784, with the peak intensity in the green regime being 0.9450 at 534.0 nm. The highest reflectance at this orientation was observed at 691.6 nm to be 1.355.

Studies conducted on the Morpho wing scales have shown that their longitudinal ridges form a diffraction structure that polarizes light.[77] The scales do not maintain the same orientation throughout the wingspan and the optical behavior of the wing section is sensitive to the position in which it is adhered to the glass slide. Therefore, this fit encompasses the global response of the Morpho wing recorded. The Jones calculus fit determined the optical axis ($\phi$) of this section of the Morpho wing to be $\phi_{Morpho}$ = 5.798°±2.237° and the extrapolated intensity maxima and minima from this fit can be found at sample stage orientations $\theta = 50.71°$ and $\theta = 5.455°$, respectively. A linearly birefringent material achieves its highest reflection intensity when its optical axis is oriented ±45° between orthogonal polarizers and its lowest reflection intensity when its optical axis is aligned with either polarizer. This confirms the birefringent behavior of the Morpho wing and its modeling as a wave retarder.

Left-handed circularly polarized light illumination with a linear analyzer (Figure 3c,d,g,h) reveals a selective color response based on the optical anisotropy of the Morpho wing. When the Morpho wing is oriented at $\theta = 135°$, the grating lines are oriented 45° relative to the direction of rotation of the electric field vector of the incident light. In turn, at a respective 135° orientation of the grating lines, this handedness is diagonally opposite to the direction of circular polarization, minimizing the interaction between the scales and the incident polarized light. This is supported by Figure 3g, which demonstrates that $\theta = 135°$ yields a higher reflectance intensity than $\theta = 45°$. This polarization-sensitive color response arises as the color observed at $\theta = 45°$ is predominantly attributed to the melanin pigments within the scales, while at $\theta = 135°$, micro- and nanostructure interference renders the color presented by the Morpho wing. These color differences are further shown as a histogram in Figure 3h, with the mean normalized Red, Green, Blue (RGB) color channel and grayscale values obtained for the images shown in Figure 3c (darker histograms) and 3d (lighter histograms).

## 2.2. Morpho-Enhanced Polarized Light Microscopy (MorE-PoL) for characterization of human breast cancer tissue

We demonstrate our MorE-PoL imaging platform with collagen-dense and collagen-sparse human breast tumor tissue sections (see section 4. Methods). Unstained formalin-fixed paraffin-embedded (FFPE) tissue sections were imaged with and without the Morpho wing between crossed linear polarizers undergoing 15° increments of sample stage rotation ($\theta = 0°\text{-}180°$) for observation of intensity variations due to optical anisotropy. For MorE-PoL imaging, the paraffin-embedding is advantageous, as it incorporates an optical diffuser in tissue-free regions of the section, removing unwanted birefringent background signal. The studied tissue sections were then deparaffinized, stained with H&E and imaged in transmission with the same polarized light microscope. The H&E-stained sections were then imaged with SHG microscopy to probe for collagen content (see section 4. Methods). The resulting images of the collagen-dense and collagen-sparse tumor tissue sections are shown in Figures 4 and 5, respectively. From the optical behavior and pathologist-interpreted H&E and SHG images (see section 4. Methods), Regions of Interest (ROIs) corresponding to Cases 1-4 of varying collagen density ($\delta$) and order ($R^2$), as presented in Figure 2, were identified in the collagen-dense and collagen-sparse tumor sections (see section 4. Methods). Jones calculus analysis (Equation 3) was implemented for the selected ROIs in images acquired when the unstained tumor tissue section was interfaced with the Morpho wing (M+T). Note that our choice of ROIs are examples to illustrate Cases 1-4 (Figure 2). Alternatively, larger ROIs could be selected for initial comparison of collagen density and order between tissues derived from different patients.

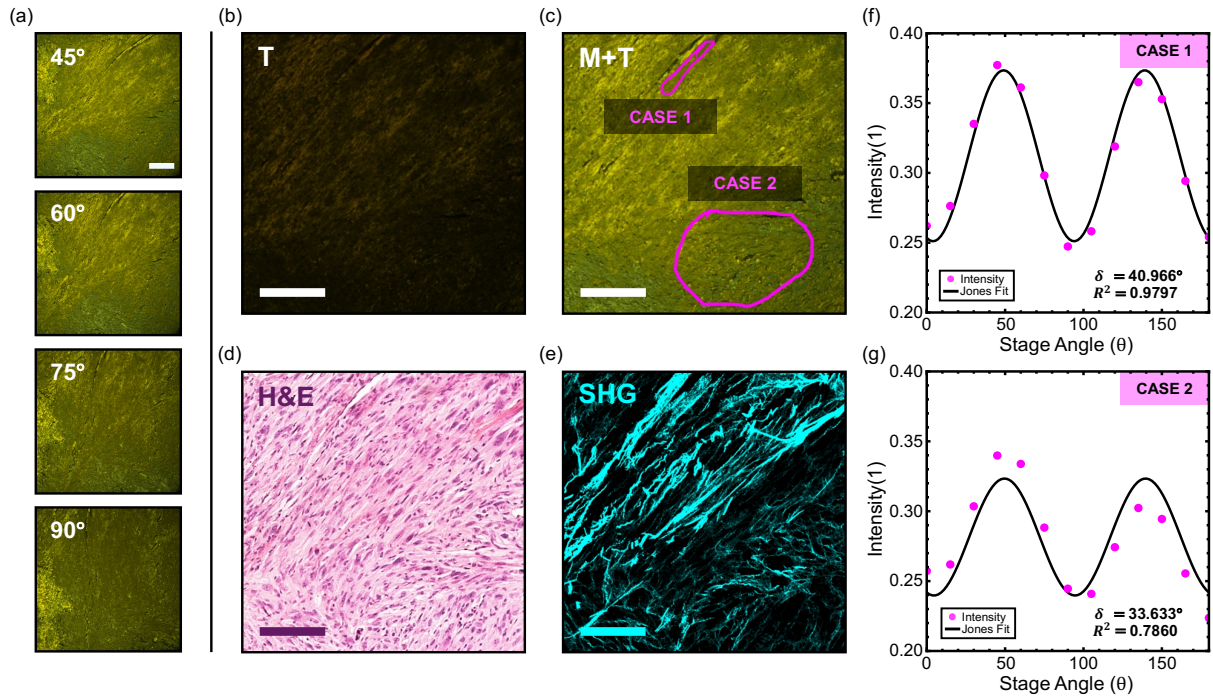

*Figure 4*: *Collagen-dense histological tumor section fiber arrangement Case 1 (ordered) and Case 2 (disordered) classification. (a) MorE-PoL images of section between crossed polarizers in reflectance for $\theta = 45° - 90°$. (b) Unstained, FFPE tumor section imaged between crossed linear polarizers. (c) Unstained, FFPE tumor section interfaced with the Morpho butterfly wing section imaged between crossed polarizers, ROIs representing Case 1 and Case 2 were encircled in FIJI. Section is positioned at $\theta=45°$. (d) Deparaffinized histological tumor section stained with H&E imaged in transmission. (e) SHG micrograph of deparaffinized H&E-stained section. Scale bars for a-e are 50 μm. (f,g) Jones calculus fit of average grayscale intensity within encircled ROI at 15° increments of microscope stage rotation from 0°-180° for Case 1 (part f) and Case 2 (part g). Parameters ϕ and δ were optimized for the Jones fit equation (Equation 3) in the MATLAB Curve Fitter tool (see section 4. Methods) to produce the curves for Case 1 (part f) and Case 2 (part g). Curves were overlaid with plots of the mean ROI pixel intensity at each stage orientation.*

**Figure 4** provides a quantitative analysis of the microstructural properties in the studied collagen-dense tissue section for two separate ROIs representing examples of Case 1 (dense, ordered) and Case 2 (dense, disordered) described in Figure 2. Figure 4a shows images of the M+T configuration at varying orientation angles $\theta$, where spatial variations in intensity with $\theta$ indicate regions of high optical anisotropy within the sample. Due to the weak natural birefringence of the tissue, these features lie below detection thresholds when the Morpho wing is not present (Figure 4b) and become visible only due to the selective enhancement of the birefringent regions of the tissue with the Morpho wing (Figure 4c). Corresponding H&E and SHG images of the same sample are shown in Figure 4d,e. The studied Case 1 and 2 ROIs, which can be selected and run through the MorE-PoL image analysis code as a free-hand drawing (see section 4. Methods), were analyzed with our Jones calculus model (Equation 3, Figure 4f,g). Jones calculus analysis resulted in corresponding $\delta$ values for the ROI representing Case 1 ($\delta_1 = 40.966°\pm 4.343°$) and Case 2 ($\delta_2 = 33.633°\pm 12.743°$), respectively. As predicted by our model, Jones calculus analysis resulted in a higher $R^2$ value for Case 1 ($R_1^2 = 0.9797$) than Case 2 ($R_2^2 = 0.7860$), indicating a comparatively higher degree of ordered collagen alignment for Case 1.

Next, we compared our MorE-PoL analysis of the studied Case 1 and Case 2 ROIs to two existing image analysis tools used to evaluate SHG micrographs: OrientationJ (an ImageJ plug-in) and CT-FIRE (see section 4. Methods and Supporting Information). OrientationJ was used to estimate the local orientation of the fibers as well as the anisotropy (coherency) and feature presence (energy) for each pixel in the image. The OrientationJ measure tool produced higher coherency and energy values for Case 1 (0.262, 462.29) than Case 2 (0.130, 176.23). CT-FIRE analysis over the same ROIs produced higher fiber volume fraction (total fiber area/ROI area) and lower angle standard deviation for Case 1 (0.6603, 53.11) than Case 2 (0.3501, 54.98). These results are in line with our MorE-PoL analysis, where Jones calculus parameters $\delta$ and $R^2$ quantify that the studied Case 1 ROI exhibits a higher degree of collagen density and organization than the studied Case 2 ROI.

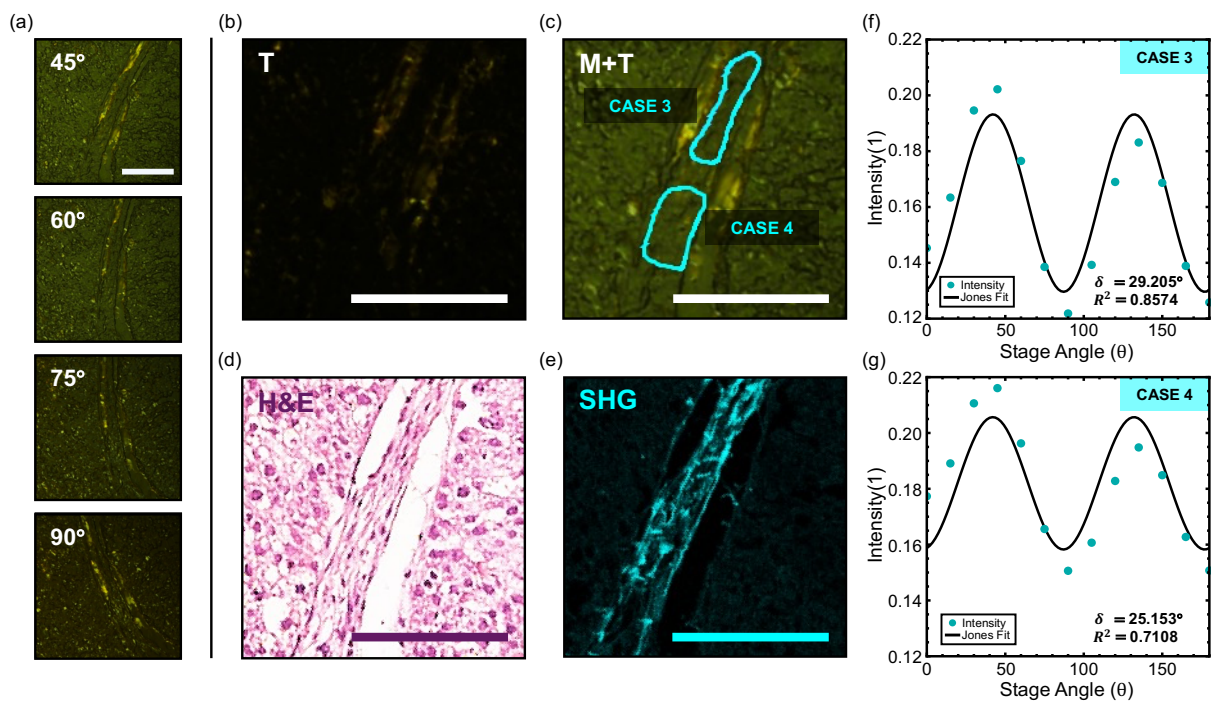

*Figure 5*: Collagen-sparse histological tumor section fiber arrangement characterization Case 3 and Case 4 classification. (a) MorE-PoL images of section between crossed polarizers in reflectance for $\theta = 45° - 90°$. (b) Unstained, FFPE tumor section imaged between crossed linear polarizers. (c) Unstained, FFPE tumor section interfaced with the Morpho butterfly wing section imaged between crossed polarizers, ROIs representing Case 3 and Case 4 are encircled in FIJI. Section is positioned at $\theta=45°$. (d) Deparaffinized histological tumor section stained with H&E imaged in transmission. (e) SHG micrograph of H&E-stained section. Scale bars for a-e are 50 µm. (f,g) Jones fit of average grayscale intensity within encircled ROI at 15° increments of microscope stage rotation from 0°-180° for Case 3 (part e) and Case 4 (part f). Parameters $\phi$ and $\delta$ were optimized for the Jones fit equation (Equation 3) in the MATLAB Curve Fitter tool (see section 4. Methods) to produce the curves for Case 3 (part f) and Case 4 (part g). Curves were overlaid with plots of the mean ROI pixel intensity at each stage orientation.

**Figure 5** quantitatively analyzes the microstructural properties of the studied collagen-sparse tissue section for two separate ROIs representing examples of Case 3 (sparse, ordered) and Case 4 (sparse, disordered), demonstrating the Morpho+Tissue configurations at varying orientation angles $\theta$ (Figure 5a) and an example orientation without and with the Morpho wing (Figure 5b,c) alongside corresponding H&E and SHG images (Figure 5d,e), respectively. Two regions of interest along the fibrillar strip were imaged (Figure 5c). Our Jones calculus analysis (Equation 3, Figure 5f,g) arrived at $\delta$ values for Case 3 ($\delta_3 = 29.205° \pm 8.583°$) and Case 4 ($\delta_4 = 25.153° \pm 11.4935°$), which lie in a similar range. Note that the $\delta$ values identified in the studied collagen-sparse Case 3 and 4 ROIs are significantly lower than those found for the studied collagen-dense Cases 1 and 2 (Figure 4), matching our model presented in Figure 2.

The $R^2$ value for this data decreased from Case 3 to Case 4 ($R_3^2 = 0.8574$, $R_4^2 = 0.7108$), indicating a higher degree of ordered collagen arrangement in the studied Case 3 ROI vs. the studied Case 4 ROI.

Image analysis with OrientationJ and CT-FIRE was applied to Cases 3 and 4 analogously to Cases 1 and 2 (see section 4. Methods). The OrientationJ measure tool produced higher coherency and energy values for Case 3 (0.314, 313.86) than Case 4 (0.162, 245.53). CT-FIRE analysis over the same ROIs produced higher fiber volume fraction and lower angle standard deviation for Case 3 (0.7950, 52.25) than Case 4 (0.4805, 53.57). These results are in line with our MorE-PoL analysis, which predicts a higher degree of collagen density and ordered arrangement for the studied Case 3 than the Case 4 ROI.

OrientationJ quantified higher coherency for Cases 1 and 3 (0.262, 0.314), which were characterized by better alignment than Cases 2 and 4 (0.130, 0.162), the cases that were characterized as more disordered. Further, OrientationJ identified higher energy for Cases 1 and 3 (462.29, 313.86) than Cases 2 and 4 (176.23, 245.53). Similarly, CT-FIRE arrived at lower angle standard deviation for Cases 1 and 3 (53.11, 52.25) than Cases 2 and 4 (54.98, 53.57). The volume fraction of fibers (i.e. fiber density) was also larger for Cases 1 and 3 (0.6603, 0.7950) than Cases 2 and 4 (0.3501, 0.4805).

We can observe that parameters quantified with OrientationJ and CT-FIRE demonstrate the differences in fiber density and anisotropy we expect for the diagnostic cases when compared within the same collagen-dense (Figure 4) or collagen-sparse (Figure 5) imaged section. However, analysis with OrientationJ and CT-FIRE faces challenges when attempting to compare parameters across Cases 1-4, i.e. ROIs from the studied collagen-dense and collagen-sparse tissue sections. In contrast, our MorE-PoL analysis enables consistent comparison of $R^2$ values and $\delta$ values across different samples.

OrientationJ and CT-FIRE both aim to characterize the orientation and presence of fibers through different mathematical evaluations. OrientationJ evaluates the structure tensor in a local neighborhood to discern the orientation and isotropic properties within a region of interest.[83] CT-FIRE functions to denoise an image and enhance fiber edge features with the fast discrete curvelet transform and a fiber extraction (FIRE) algorithm to extract individual fibers.[84] As these softwares are applied to 2D images, the computation operates under the assumption that the fibers are predominantly positioned on the imaging plane, which neglects the contribution of out-of-plane fibers. Additionally, smaller ROIs invite more background noise influence in calculating the parameters for both techniques. As OrientationJ and CT-FIRE both rely on fine elements to determine feature and fiber properties, this can pose a problem when trying to isolate and accurately characterize small biological features. Large brightness gradients corresponding to non-uniform collagen distribution adversely affects the calculation of coherency in OrientationJ as well as fiber thickness and length in CT-FIRE, resulting in an inaccurate interpretation of anisotropy.[84,85] CT-FIRE also poses challenges in adequately calculating the thickness and length of fibers that are close to each other and have joints with other fibers.[84] Thus, attempting to accurately characterize collagen disorder in these instances may prove difficult.

When comparing MorE-PoL to results from OrientationJ and CT-FIRE, the highest coherency values should follow the highest $R^2$ values, and the same is expected for high $\delta$ values and high energy values, yet this is not the case when comparing values across samples. This observation can be attributed to some of the drawbacks of interpreting over an SHG micrograph, as signal enhancement occurs primarily on the surface and only on surfaces perpendicular to a fiber's optical axis.[86] SHG polarization is heightened when the laser is parallel to the surface of the fibers, and the SHG radiates from the shell rather than the bulk of the collagen fiber.[86] Orientation sensitivity can be mitigated by acquiring images over several angles of rotation, which can be difficult to accomplish with commercial systems. The benefit of our MorE-PoL technique is that the polarized light propagates throughout the tissue volume, thus the reflected light characterizes the behavior of all the fibers within the section, rather than the surface-level fibers seen in the SHG image. This can explain the differences we observe between the OrientationJ and CT-FIRE SHG micrograph analysis and our MorE-PoL analysis.

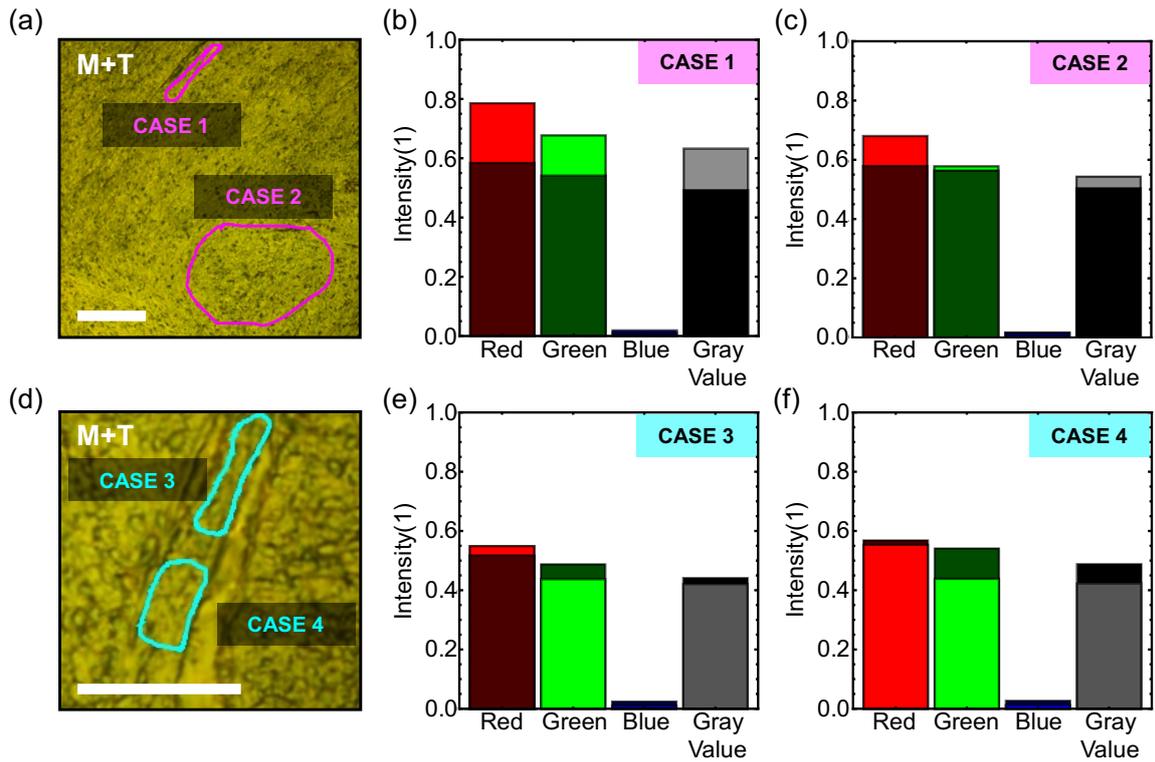

***Figure 6:*** *MorE-PoL of the studied collagen-dense and collagen-sparse histological tumor sections for circularly polarized light illumination with a linear analyzer. (a) MorE-PoL images of collagen-dense section for stage orientation θ=45°. ROIs representing Cases 1 and 2. (b,c) Average normalized RGB color channel and grayscale pixel intensity for Case 1 (part b) and Case 2 (part c) for 45° (darker histograms) and 135° (lighter histograms) stage rotation. (d) MorE-PoL images of collagen-sparse section for stage orientation θ=45°. ROIs representing Cases 3 and 4. (e,f) Average normalized RGB color channel and grayscale pixel intensity for Case 3 (part e) and Case 4 (part f) for 45° (darker histograms) and 135° (lighter histograms) stage rotation. Scale bars are 50 μm. Pixel values were calculated in MATLAB.*

**Figure 6**a,d displays images of the studied collagen-dense (Figure 6a, Case 1,2 ROIs indicated, $\theta = 45°$) and collagen-sparse (Figure 6d, Case 3,4 ROIs indicated, $\theta = 45°$) tissue sections, for left-handed circularly polarized light illumination and a linear analyzer. The mean normalized RGB color channel and grayscale value was interpreted over each ROI and angle of orientation to document the change in colorimetric response. Different intensity values for $\theta = 45°$ and $\theta = 135°$ stage orientations are demonstrated for Cases 1 and 2 (Figure 6b,c) and Cases 3 and 4 (Figure 6e,f). The different orientations are represented by either the true red, green, blue, or black RGB color ($\theta = 135°$) or the darkened red, green, blue, or black RGB color ($\theta = 45°$). Pixel RGB and grayscale values were calculated in MATLAB. The change in RGB values at each orientation produces uniquely different color responses that can be detected by the observer. These different colorimetric properties can also be observed when comparing the collagen-dense and collagen-sparse tumor sections, in which we see strong RGB channel

and grayscale intensities in the high collagen tumor section, and the greatest color contrast for Case 1. Thus, MorE-PoL imaging with circularly polarized illumination and a linear analyzer provides a stain-free method to visualize the morphological and microstructural properties of a studied tissue section, which can aid with ROI identification for further analysis of the tissue microstructure with our Jones calculus model and linearly polarized light illumination.

**3. Conclusion**

In conclusion, we introduced Morpho-Enhanced Polarized Light Microscopy (MorE-PoL), a label- and contact-free imaging platform to quantitatively assess the microstructural properties of fibrous biological tissue sections by leveraging the intrinsic optical anisotropy of the Morpho butterfly wing. MorE-PoL presents significant advantages for improved precision and accessibility of tissue-microstructure imaging: (i) It can be directly implemented in a commercial polarized light microscope and does not necessitate complex and costly equipment. (ii) It circumvents laborious and artifact-prone staining procedures. (iii) The Morpho wing is derived from nature and does not require micro- or nanoscale fabrication techniques. (iv) The Morpho wing can be reused repeatedly, due to the contact-free methodology. (v) MorE-PoL is neither disease- nor organ-specific and widely applicable to versatile fibrotic tissues. These advantages can enable facile implementation of MorE-PoL, including in under-resourced settings. We achieved quantitative analysis of biological tissue microstructure by introducing a mathematical image analysis model based on Jones calculus. As a representative example system, we studied collagen-dense and collagen-sparse human breast cancer tissue sections.

Our Jones calculus model determines two parameters for tissue fiber characterization, $\delta$ which relates to fiber density, and $R^2$ that corresponds to fiber anisotropy. This model arrived at the highest $\delta$ and $R^2$ for Case 1 (40.966°± 4.343°, 0.9797), and the lowest $\delta$ and $R^2$ for Case 4 (25.153°±11.4935°, 0.7108) which falls in line with the characteristic properties for each case (high density and low disorder vs. low density and high disorder). The same applies for Cases 2 (high density and high disorder) and Case 3 (high density and low disorder) falling in line with their respective $\delta$ (°33.633° ± 12.743°, 29.205°± 8.583°) and $R^2$ (0.7860, 0.8574) values.

It is worth noting that collagen is not the only fibrous tissue or component that exhibits form birefringence. Therefore, our Jones calculus model (Equation 3) is not limited to collagen. This work provided examples using the intentionally selected regions identified to predominantly exhibit fibrillar collagen from pathologist assessment of the SHG micrographs and H&E stains. However, the presented platform and method can be applied to investigate the optical behavior of other extracellular fibers and various material microstructures. The inventive scheme provides a low-cost, contact- and label-free imaging approach, employing a readily available natural material, for the evaluation of both natural and synthetic forms of material microstructure. The presented application to biological tissues enables the imaging and quantitative assessment of changes in tissue microstructure, and an improved understanding of its role in the origin and progression of a broad range of fibrotic diseases.

## 4. Experimental Section/Methods

*Morpho Butterfly Wing Sample Preparation and Characterization*: The *M. menelaus* butterfly studied in this work was purchased from BioquipBugs and sourced from Peru. Image of the Morpho butterfly (Figure 1b) was taken with an iPhone 8 camera. Scanning electron micrograph (SEM) of the Morpho wing scales was acquired with an FEI Apreo LoVac SEM. Spectra of the Morpho wing at the recorded orientations producing its minimum (90°) and maximum (150°) intensities under linearly- and circularly polarized light illumination with a linear analyzer were acquired with an IsoPlane 320A Spectrometer (Figure 3).

*Biological Tissue Sample Preparation:* Patient derived xenograft (PDX) models of a range of treatment-naïve triple-negative breast cancer (TNBC) subtypes were previously established by Dr. Helen Piwnica-Worms (MD Anderson).[87] Two PDX models, named PA14-13 and PIM025, were selected based on their stiffness (respectively, dense vs. sparse collagen content) determined by SHG microscopy, and expression of the Epithelial-mesenchymal transition (EMT) transcription factor Twist1. 500,000 PDX cells were implanted into the fourth mammary fat pads of immunodeficient mice, corresponding to Institutional Animal Care and Use Committee (IACUC) guidelines for animal care. When tumors reached 1000 mm$^3$, tumor tissues were collected, fixed with 4% Paraformaldehyde at 4°C for 48h and embedded in paraffin. These fixed paraffin embedded tissues were sectioned at 3 μm thickness.

*Optical Characterization:* Samples were imaged with a Nikon ECLIPSE LV100ND polarized light microscope with a Nikon LV-HL 50W 12V LONGLIFE halogen lamp and 20x objective. For polarized light microscopy a C-SP simple polarizer introduced polarized light incident onto the sample and an analyzer with a 360° rotary dial was inserted into the light path. For crossed polarized light microscopy, the input polarizer was positioned along the horizontal and the analyzer was positioned vertically. For circularly polarized light imaging left-handed circularly polarized light was incident onto the sample, and the analyzer was oriented along the vertical for Cases 1-4 (Figure 6), and along the horizontal for imaging the Morpho wing by itself (Figure 3). Incident lamp intensity was consistent across all images for Cases 1-4. Intensity was reduced for imaging the Morpho wing alone under circularly polarized light illumination to prevent camera saturation. Images were acquired in the NIS Elements Basic Research software and saved as 24-bit uncompressed TIFF files. The pixel size was set to 0.435 μm.

Spectra of the Morpho wing were collected under the aforementioned imaging conditions with the Princeton Instruments IsoPlane 160 Spectrometer and acquired in LightField 64-bit Data Acquisition Software. Acquisition was conducted under the "Full Sensor, Binned" condition (Bin W = 1, Bin H = 400). The exposure time was set to 10000 ms, center wavelength was 500 nm, slit width was 60 μm, and the grating was 300g/mm. Spectra were normalized by dividing the Morpho reflectance by the reflectance spectra of a ThorLabs Protected Silver Mirror (PF10-03-P01 ⌀1") at the same incident intensity. The MATLAB smoothdata function was used to smooth the normalized spectra with a Gaussian-weighted moving average filter over an 80 nm window.

*Pathologist Analysis of Biological Tissue Samples:* Tissues were stained with hematoxylin and eosin, following standard procedures. H&E photomicrographs and SHG images were provided to a board-certified pathologist, who indicated that the tissue in the Case 1 and 2 image (corresponding to PA14-13 PDX tumor model) was likely to contain more collagen than the tissue in the Case 3 and Case 4 image (from PIM025 tumors). Using QuPath software,[14] the pathologist annotated regions which were expected to have the highest fibrillar collagen content.

*Morpho-Enhanced Polarized Light Microscopy (MorE-PoL)*: A ~1"x1" section a *Morpho menelaus* butterfly wing was cut out and adhered to a glass slide with clear nail polish. A cover slip was adhered on top to prevent scales from escaping the wing surface. The slide containing the Morpho wing section was placed underneath the slide with the histological tumor section and placed on the rotating stage of the polarized light microscope. This arrangement was imaged between crossed linear polarizers and the stage was rotated counterclockwise from 0°-180° in 15° increments.

*Region of Interest (ROI) Selection*: ROI's corresponding to Cases 1-4 in the collagen-dense and collagen-sparse tumor sections were identified with freehand selections in FIJI (a distribution of ImageJ) and saved as regions of interest with FIJI's ROI manager. The XY coordinates of the ROI were saved as comma-separated value (CSV) files. These XY-coordinates and M+T polarized light microscopy images for orientations 0°-180° were imported into MATLAB. The detectSURFFeatures and estgeoform2d functions were used to detect the geometric transformation between images acquired between successive 15° increments (for further information reference MATLAB's "Find Image Rotation and Scale Using Automated Feature Matching" documentation). The transformation matrix was first applied to the XY-coordinates of the original images isolated in FIJI ($\theta=45°$) to be adapted to the 30° image. The transformation matrices for the preceding orientations (15° − 0°) and images following the 45° image (60° − 180°) were applied to the appropriate coordinates. The mean grayscale intensity of pixels within the ROI was calculated for each orientation to be implemented for the Jones calculus analysis.

*Jones Calculus Analysis*: For each image of the wing section, the pixel intensity was averaged and normalized by maximum pixel intensity (255) in MATLAB. The average intensity for each orientation was plotted and fitted to the Jones calculus model with MATLAB's built-in nonlinear least squares function in the Curve-Fitting tool. The tool was implemented to estimate $\phi$ and $\delta$ values that produce the best fit of the raw data. The non-linear least squares method and trust-region algorithm were selected. For the simulated Cases 1-4 in Figure 2, the constraints $\phi \in [0°, 180°]$ and $\delta \in [0°, 180°]$ were applied to the algorithm. The $R^2$ value of the fit is generated by the curve-fitting tool. $\delta = 90°$ and $\delta = 45°$ was inserted into the Jones intensity profile for $\uparrow \delta$ and $\downarrow \delta$ cases, respectively. For $\uparrow R^2$ $\phi = 0$ was plugged into the intensity profile. To generate a profile with $\downarrow R^2$, an array of 13 random $\phi$ values for $\phi \in [0°, 45°]$ was generated with the MATLAB randi function. This array was then inserted into the Jones fit equation (Equation 3) to produce intensities for each 15° of rotation. The resulting values were fit to the Jones fit equation in the Curve-Fitting tool to produce the curves in Figure 2. For the experimentally measured Cases 1-4 from MorE-PoL imaging (Figures 4,5), the mean grayscale intensity at each orientation were inserted into the Jones intensity fit in the Curve-Fitting tool with the constraints $\phi \in [-180°, 180°]$ and $\delta \in [0°, 180°]$ applied to the algorithm.

*Second Harmonic Generation Microscopy:* The second-harmonic generation imaging was performed on an upright Leica SP8 microscope with a resonant scanner and hybrid non-descanned detectors. Ti-Sapphire femtosecond pulsed laser (Chameleon Ultra II, Coherent Inc.) was tuned to 855 nm and the beam was focused on the sample with an HC PL APO CS 10x/0.40 dry objective. The light was routed to the detectors with 560 nm, 495 nm, and 605 nm long-pass dichroic mirrors. The SHG signal was recorded with a 425/26 nm bandpass filter, the autofluorescence was recorded with a 525/50 nm bandpass filter. The pixel size was set to 0.311 µm, and 32x line averaging was used to improve the signal-to-noise ratio. Data were digitized in an 8-bit mode. The sample navigator software module was used to create autofocus support points and individual fields of view were tiled and stitched.

*CT-FIRE*: For the CT-FIRE analysis, the ROIs were flattened onto the SHG micrographs in FIJI and the resulting images were uploaded to CT-FIRE. The ROIs were retraced and saved with the ROI manager. These ROIs were then applied to the original SHG micrographs in grayscale and analyzed with the CTF ROI analyzer. The image resolution was set to 300 dpi, minimum fiber length to 3 pixels, max fiber width to 800 pixels and histogram bins to 15. The pixel area of each ROI was measured in CT-FIRE. The total area of detected fibers within the ROI was calculated by taking the sum of the multiplied pixel length and pixel width of each fiber. The volume fraction was determined by dividing this value by the measured pixel area of the entire ROI.

**Acknowledgements**

The authors thank Zaid Haddadin, Samantha Bordy, Omonigho Aisagbonhi, Sabine Faulhaber, Zbigniew Mikulski and Kenneth Kim for helpful discussions. L.V.P. and P.K. gratefully acknowledge funding from the Beckman Young Investigator award by the Arnold and Mabel Beckman Foundation (Project Number: 30155266). P.K. additionally acknowledges funding from the Optica Amplify Scholarship, the Optica Women's Scholarship and the Selected Professions Fellowship from the American Association of University Women (AAUW). This work was also supported by grants from NCI (R01CA174869, RO1CA262794, R01CA268179, and R01CA236386) and Krueger v. Wyeth research award to J.Y. A.M.F. is supported by TRDRP Postdoctoral Award T32FT4922.

**Table of Contents Entry**

**Leveraging Optical Anisotropy of the Morpho Butterfly Wing for Quantitative, Stain-Free, and Contact-Free Assessment of Biological Tissue Microstructures**

*Paula Kirya, Aida Mestre-Farrera, Jing Yang, Lisa V. Poulikakos\**

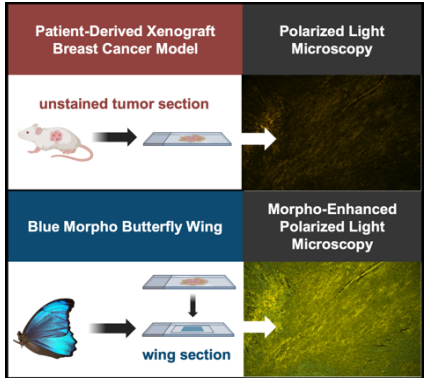

*We leverage the intrinsic optical anisotropy of the Morpho butterfly wing to introduce Morpho-Enhanced Polarized Light Microscopy (MorE-PoL), a stain- and contact-free imaging methodology which quantitatively assesses the microstructural properties of fibrous biological tissues. We demonstrate the potential of MorE-PoL with patient-derived xenograft breast cancer tissues, where we quantify the density and organization of fibrillar collagen in the tumor microenvironment.*